\documentclass[doublecol,figures]{epl2}
\usepackage{amssymb,epsfig,setspace,graphicx}  
\usepackage{amsmath}

\begin{document}
\hyphenation{ge-ne-ra-tes dif-fe-rent}
\hyphenation{me-di-um  as-su-ming pri-mi-ti-ve pe-ri-o-di-ci-ty}
\hyphenation{mul-ti-p-le-sca-t-te-ri-ng i-te-ra-ti-ng e-q-ua-ti-on}
\hyphenation{wa-ves di-men-si-o-nal ge-ne-ral the-o-ry sca-t-te-ri-ng}
\hyphenation{di-f-fe-r-ent tra-je-c-to-ries e-le-c-tro-ma-g-ne-tic pho-to-nic}
\hyphenation{Ray-le-i-gh di-n-ger}

\title{On the spectrum of a class of quantum models}

\author{
Alexander Moroz 
} 
\institute{Wave-scattering.com}


\abstract{
The spectrum of any quantum model 
which  eigenvalue equation reduces 
to a three-term recurrence, such as a displaced harmonic oscillator,
the Jaynes-Cummings model, 
the Rabi model, and a generalized Rabi model, can be determined 
as zeros of a corresponding transcendental function 
$F(x)$. The latter can be analytically determined as an infinite series 
defined solely in terms of the recurrence coefficients.
}

%

\maketitle

%

\section{Introduction}
\label{sc:intr}
The present work deals with a class ${\cal R}$ of 
quantum models described by
a Hamiltonian $\hat{H}$ characterized in that the eigenvalue
equation  
\begin{equation}
\hat{H}\phi=E\phi
\label{eme}
\end{equation}
in the Bargmann Hilbert space 
of analytical functions ${\cal B}$ \cite{Schw,Brg} reduces to 
a {\em three-term difference equation}
\begin{equation}
c_{n+1} + a_n c_n + b_n c_{n-1}=0 \hspace*{1.8cm}  (n\ge 0).
\label{3trg}
\end{equation}
Here $\{c_n\}_{n=0}^\infty$ are the sought expansion coefficients of 
a physical state described by an entire function 
\begin{equation}
\phi(z)=\sum_{n=0}^\infty c_n z^n.
\label{pss}
\end{equation}
The recurrence coefficients $a_n$ and $b_n$ are functions 
of model parameters and we require that $b_n\neq 0$.
In what follows, models of the class ${\cal R}$ 
are characterized by the recurrence coefficients
having an asymptotic power-like dependence 
\begin{equation}
a_n\sim a n^{\delta},~~~~ b_n\sim b n^{\upsilon}~~~~ (n\rightarrow\infty),
\label{rcd}
\end{equation}
wherein $2\delta>\upsilon$ and $\tau=\delta-\upsilon\geq 1/2$, 
which guarantees $\lim_{n\rightarrow\infty} b_n/a_n=0$. 

The conventional boson
annihilation and creation operators $\hat{a}$ and $\hat{a}^\dagger$ 
satisfying commutation relation 
$[\hat{a},\hat{a}^{\dagger}] = 1$ are represented in the Bargmann 
Hilbert space ${\cal B}$ as \cite{Schw,Brg}
\begin{equation}
\hat{a} \rightarrow \frac{\partial}{\partial z}, 
\hspace*{1.2cm} \hat{a}^\dagger \rightarrow z.
\end{equation}
Therefore, the class ${\cal R}$ comprises among others
single-mode boson models represented by
\begin{equation}
\hat{H}= A_1 \hat{a}^\dagger \hat{a}  
 +  A_2 \hat{a}^\dagger + A_3 \hat{a} + A_4,
\label{mcr}
\end{equation}
where $A_j$ are in general matrix coefficients.
Indeed, on comparing the same powers of $z$ on both sides of 
(\ref{eme}), the consecutive $\hat{a}^\dagger \hat{a}$,
$\hat{a}^\dagger$, and $\hat{a}$ terms in (\ref{mcr}) lead
to $nc_n$, $c_{n-1}$, and $(n+1)c_{n+1}$ 
terms in the recurrence (\ref{3trg}), respectively. 
Then the eigenvalue
equation (\ref{eme}) reduces to a coupled system 
of ordinary differential equations \cite{Schw,Ks,KL}.
The model Hamiltonian (\ref{mcr}) encompasses many prominent 
examples, such as a displaced harmonic oscillator \cite{Schw}, 
the Rabi model \cite{Schw,Rb}, the 
Jaynes and Cummings (JC) model proposed as an approximation to
the Rabi model \cite{JC}, and a generalized Rabi model \cite{Br}.
The Rabi model describes the simplest interaction between a 
cavity mode with a bare frequency 
$\omega$ and a two-level system 
with a bare resonance frequency $\omega_0$. 
The model is characterized by 
the Hamiltonian \cite{Schw,Rb,Br}, 
\begin{equation}
\hat{H}_R =
\hbar \omega \hat{a}^\dagger \hat{a}  
 +  \lambda\sigma_1 (\hat{a}^\dagger + \hat{a}) + \mu \sigma_3,
\label{rabih}
\end{equation}
where 
$\lambda$ is a coupling constant, and $\mu=\hbar \omega_0/2$. 
In what follows we assume the standard representation
of the Pauli matrices $\sigma_j$ and set the Planck constant $\hbar=1$.  
\begin{figure}
\onefigure[scale=0.7]{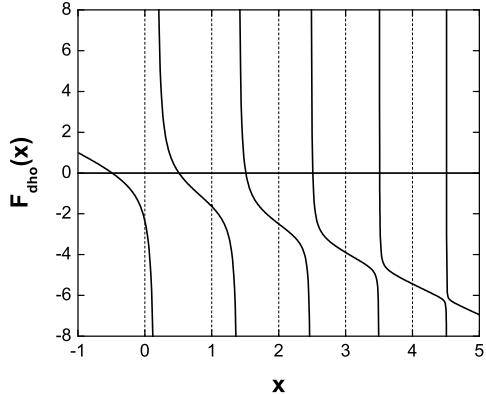}
\caption{$F_{dho}$ as a function of $x=E/\omega$ 
for $\kappa=\lambda/\omega=0.7$, $\Delta=\mu/\omega=0$,
in the case when the spectrum of the Rabi model reduces to that
of a displaced harmonic oscillator. In agreement 
with the exact analytic formula (\ref{alqc}), the zeros of $F_{dho}(x)$ 
satisfy $x_l=l-0.49$.}
\label{fgdho}
\end{figure}

Our main results is that the spectrum of any
quantum model from ${\cal R}$ can be obtained as zeros of 
a transcendental function defined by infinite series 
\begin{equation}
F(x)\equiv a_0 + \sum_{k=1}^\infty \rho_1\rho_2\ldots\rho_k
\label{fdf}
\end{equation}
solely in terms of the coefficients of 
the three-term recurrence in eq. (\ref{3trg}) \cite{AMcm}:
\begin{eqnarray}
\rho_1 &=& -\frac{b_1}{a_1},
\hspace*{0.3cm}
\rho_l=u_l-1,
\hspace*{0.3cm}
u_1=1,
\nonumber\\
u_l &=& \frac{1}{1-u_{l-1}b_l/(a_la_{l-1})},
\hspace*{0.3cm}
l\geq 2.
\label{eth}
\end{eqnarray}
The function $F(x)$ is different from the $G$-functions of 
Braak \cite{Br}. The latter have only been obtained 
in the case of the Rabi model by making explicit use of
the parity symmetry. In contrast to
Braak's result \cite{Br}, our $F(x)$ can be 
straightforwardly defined for any model from ${\cal R}$.
The definition of $F(x)$ does not require either a discrete symmetry
or to solve the three-term difference equation (\ref{3trg}) explicitly.
All what is needed to determine $F(x)$ is an explicit knowledge of 
the recurrence coefficients $a_n$ and $b_n$. 
This brings about a straightforward numerical 
implementation \cite{AMr} and results in a
great simplification in determining the
spectrum. The proof of principle 
is demonstrated in fig. \ref{fgdho} in the case of 
exactly solvable displaced harmonic oscillator \cite{Schw}.
The ease in obtaining the spectrum is of importance 
regarding recent experimental advances in preparing 
({\em ultra})strongly interacting 
quantum systems \cite{BGA,FLM,NDH,SHu,CLO}, which can 
no longer be reliably described by the 
exactly solvable JC model \cite{JC}, 
and wherein only the full quantum Rabi model can describe the observed 
physics.

\section{Proof of the main result}
\label{sc:rs}
In earlier studies \cite{Schw,Ks,KL,Br},
largely motivated by the Frobenius method \cite{Frb}
of solving differential equations, the $n=0$ part of the 
recurrence (\ref{3trg}) was taken as an initial condition. 
Given the requirement of analyticity [$c_{-k}\equiv 0$ for $k>1$;
{\em cf.} eq. (\ref{pss})], the three-term recurrence 
(\ref{3trg}) degenerates for $n=0$ into an equation involving 
mere {\em two terms} and imposes that 
\begin{equation}
c_1/c_0=-a_0.
\label{rbc}
\end{equation} 
By solving the recurrence upwardly,
one as a rule arrives at the expansion coefficients 
of {\em singular} functions {\em outside}
the physical Bargmann Hilbert space ${\cal B}$ \cite{Schw,Ks,Frb}. 
Only at a particular discrete set of parameter values that
correspond to the spectrum of a model the 
functions would belong to ${\cal B}$ \cite{Schw,KL}.
Our approach differs from earlier studies \cite{Schw,KL,Br}
in that our focus is initially on the truly 
{\em three-term} part of the recurrence (\ref{3trg})
for $n\ge 1$. This would enable us 
to remain in ${\cal B}$
for any value of physical parameters. The remaining $n=0$ part of the 
recurrence (\ref{3trg}), 
which reduces to a relation between $c_1$ and $c_0$,
then becomes the {\em boundary
condition} for $\phi\in{\cal B}$ defined by 
the $n\ge 1$ solutions of the recurrence (\ref{3trg}).

In virtue of the Perron and Kreuser generalizations 
(Theorems 2.2 and 2.3(a) in ref. \cite{Gt}) of the Poincar\'{e} theorem 
(Theorem 2.1 in ref. \cite{Gt}), the difference 
equation (\ref{3trg}) (considered for $n\ge 1$) possesses in our case
{\em two} linearly independent solutions:
\begin{itemize}

\item (i) a {\em dominant} solution $\{d_j\}_{j=0}^\infty$ and

\item  (ii) a {\em minimal} solution $\{m_j\}_{j=0}^\infty$.
\end{itemize}

The respective solutions differ in the 
behavior $c_{n+1}/c_n$ in the limit $n\rightarrow\infty$.
The {\em minimal} solution guaranteed by 
the Perron-Kreuser theorem
(Theorem 2.3 in ref. \cite{Gt}) for models 
of the class ${\cal R}$ satisfies 
\begin{equation}
\frac{m_{n+1}}{m_n}\sim -\frac{b}{a}\frac{1}{n^\tau} \rightarrow 0
\hspace*{0.8cm} (n\rightarrow \infty)
\label{mins}
\end{equation}
in virtue of (\ref{rcd}) and $\tau \geq 1/2>0$.
On substituting the minimal solution 
for the $c_n$'s in eq. (\ref{pss}), 
$\phi(z)$ automatically becomes an {\em entire} function.
In what follows, the entire function 
generated by the minimal solution of the $n\ge 1$ part 
of (\ref{3trg}) will be denoted as $\phi_m(z)$.
Now (\ref{mins}) presumes 
[up to the terms reducing to a ${\cal O}(1)$ term
in (\ref{mins})] an asymptotic behaviour
\begin{equation}
m_n \sim k^n/[\Gamma(n+1)]^\tau
\label{minas}
\end{equation}
with $k=-b/a$.
A sufficient condition for a minimal solution to generate 
$\phi_m(z)\in{\cal B}$ can be read off
from asymptotic properties of the generalized Mittag-Leffler function
$E_{\tau,\gamma}(z)$ ({\em cf.} Sec. 3 of \cite{AMds}), 
an entire function with the
expansion coefficients $1/\Gamma(\tau n+\gamma)$, where  
$\tau$ and $\gamma$ are real constants. In brief
$\phi_m(z)\in{\cal B}$ if
the leading-order decay of $m_n$ for $n\rightarrow\infty$ is not slower 
than $\propto k^n/\Gamma(\tau n+\gamma)$, where $k$ is a constant 
and $\tau>1/2$. For $\tau=1/2$
the otherwise unrestricted constant $k$ has to satisfy $|k|</\sqrt{2}$.
Up to an irrelevant
proportionality constant,
\begin{equation}
[\Gamma(n+1)]^\tau \approx (n)^{\tau/2} (\tau^\tau)^n \Gamma(\tau n+1).
\label{gmrl}
\end{equation}
Therefore, when $\Gamma(\tau n+1)$ had been substituted by 
$[\Gamma(n+1)]^\tau$ in $E_{\tau,\gamma}(z)$,
one would have arrived at substantially the same conclusions, but with 
$k$ replaced by $k\tau^\tau$. For $\tau=1/2$ the latter yields $k/\sqrt{2}$,
and hence the sufficient condition 
$|k|<1$ for $\phi_m(z)$ to belong to ${\cal B}$ 
({\em cf.} eq. (1.4) of ref. \cite{Brg}).

Now it is important to realize that only 
the ratios of subsequent 
terms $m_{n+1}/m_{n}$ of the {\em minimal} solution 
are related to infinite continued fractions 
({\em cf.} Theorem 1.1 due to Pincherle in ref. \cite{Gt})
\begin{equation}
r_{n}= \frac{m_{n+1}}{m_{n}} 
    = \frac{-b_{n+1}}{a_{n+1}-} \frac{b_{n+2}}{a_{n+2}-}
\frac{b_{n+3}}{a_{n+3}-}\cdots
\label{rncf}
\end{equation}
Any {\em minimal} solution is, 
up to a multiplication by a constant, {\em unique} \cite{Gt}.
This has two immediate consequences. First, $\phi_m(z)$ is, 
up to a multiplication constant, {\em unique}.
Second, the ratio $r_0=m_{1}/m_0$ of the first two terms 
of a given minimal solution is {\em unambiguously} fixed.
Note that the ratio $r_0=m_1/m_0$ involves $m_0$, although it
takes into account the recurrence (\ref{3trg}) only for $n\geq 1$ \cite{Gt}.
The remaining $n=0$ part of the recurrence (\ref{3trg}) 
imposes another condition [{\em cf.} eq. (\ref{rbc})]
on the ratio $r_0=m_1/m_0$. 
The condition (\ref{rbc}) can be translated 
into the {\em boundary condition} on the logarithmic
derivative of $\phi_m(z)$,
\begin{equation}
r_{0}=\frac{\phi_m'(0)}{\phi_m(0)}=-a_0.
\label{bcc}
\end{equation}
Obviously, one has $r_0\neq -a_0$ in general. 
As a rule it is impossible to find a minimal solution 
satisfying an arbitrarily prescribed initial condition 
on the ratio $m_{1}/m_0$ of its 
first two terms. (This point was not made clear 
by either Schweber \cite{Schw} or later on by Braak \cite{Br}.)
If one defines the function 
$F=a_0+r_0$, with $r_0$ given by the continued fraction in (\ref{rncf}),
the zeros of $F$ would 
correspond to those points in a parameter space where the condition 
(\ref{bcc}) is satisfied. Then the coefficients of 
$\phi_m(z)\in{\cal B}$ satisfy the recurrence (\ref{3trg}) including $n=0$,
and $\phi_m(z)$ belongs to the spectrum. 
Analogous to the
Schr\"{o}dinger equation, the boundary condition (\ref{bcc})
enforces quantization of energy levels.

The rest of the proof follows upon an 
application of the {\em Euler theorem}, which transforms the
continued fraction defining $r_{0}$ in eq. (\ref{rncf}) 
into an infinite series
({\em cf.} eqs. (4.4)-(4.5) on p. 43 of \cite{Gt}
forming the basis of the ``{\em third}" method of Gautschi \cite{Gt}
at the special case of his $N=0$
that is also used in our numerical implementation \cite{AMr}).
The latter leads directly to eqs. 
(\ref{fdf}) and (\ref{eth}) for $F$ \cite{AMcm}.
The series in (\ref{fdf}) converges whenever the
continued fraction in (\ref{rncf}) converges for $n=0$, 
which occurs if and only if $\phi_m(0)=m_0\ne 0$ 
({\em cf.} Theorem 1.1 due to Pincherle in ref. \cite{Gt}).

\section{Illustration of some basic properties}
\label{sc:ils}
In the case of a displaced harmonic oscillator, which 
is the special case
of $\hat{H}_R$ in (\ref{rabih}) for $\mu=0$,
the recurrence (\ref{3trg}) becomes ({\em cf.} 
eq. (A.17) of ref. \cite{Schw})
\begin{equation}
c_{n+1} + \frac{n-x}{(n+1)\kappa}\, c_{n} + \frac{1}{n+1}\, c_{n-1}=0,
\label{sa17}
\end{equation}
where dimensionless parameter $\kappa=\lambda/\omega$ reflects
the coupling strength and
$x=\epsilon=E/\omega$ is a dimensionless energy parameter.
Energy levels satisfy \cite{Schw}
\begin{equation}
\epsilon_l = E_l/\omega =l-\kappa^2,
\label{alqc}
\end{equation}
where $l\in\mathbb{N}$ is a {\em nonnegative} integer (including zero). 
As demonstrated in fig. \ref{fgdho}, $F_{dho}(x)$
defined by eqs. (\ref{fdf}) and (\ref{eth}) 
displays for $\omega=1$ and $\kappa=0.7$ 
a series of {\em discontinuous} branches 
extending {\em monotonically} between $-\infty$ and $+\infty$,
which intersect the $x$-axis at $x_l=l-0.49$,
$l=0,1,\ldots$. The zeros correspond exactly to the position of 
the energy levels (\ref{alqc}).

In the present example one can also explicitly illustrate 
that the difference equation (\ref{3trg}) with
the condition (\ref{rbc}) imposed as an initial condition 
of the recurrence, and then solved upwardly, always possesses a solution. 
Obviously, such a solution would be 
generally a {\em dominant} solution.
Consequently, the function $\phi(z)$ defined by power 
series expansion (\ref{pss})
with the expansion coefficients $\{d_j\}_{j=0}^\infty$ would 
exhibit (typically branch-cut \cite{Schw}) {\em singularities} in the 
$z$-complex plane and would not belong to ${\cal B}$.
Indeed, the recurrence (\ref{sa17}) is solved for $n\ge 0$ with
\begin{equation}
c_n=\kappa^{\alpha-n} L_{n}^{(\alpha-n)}(\kappa^2),
\label{cns}
\end{equation}
where $\alpha= x +\kappa^2$ and $L_{n}^{\beta}$ 
are associated Laguerre polynomials \cite{Lgp}
(note different sign of $\kappa$ compared to eq. 
(2.16) of Schweber \cite{Schw}).
The substitution (\ref{cns}) transforms (\ref{sa17}) 
into a $3$-point rule
that is identically satisfied by the associated 
Laguerre polynomials \cite{Lgp}.
The Rodrigues formula \cite{Lgp}
\begin{equation}
    L_n^{(\alpha)}(z)= {z^{-\alpha} e^z \over n!}{d^n \over dz^n} 
                  \left(e^{-z} z^{n+\alpha}\right)
\end{equation}
implies $L_{0}^{(\alpha)}(z) = 1$, $L_{1}^{(\alpha)}(z)= -z + 1 + \alpha$,
and in virtue of (\ref{cns})
\begin{eqnarray}
c_0 &=& \kappa^{\alpha} L_{0}^{(\alpha)}(\kappa^2) =\kappa^{\alpha},
\nonumber\\
c_1 &=& \kappa^{\alpha-1} L_{1}^{(\alpha-1)}(\kappa^2) = \kappa^{\alpha-1} x.
\end{eqnarray}
Thus $c_1/c_0= x/\kappa$, which is exactly the $n=0$ part of (\ref{sa17}).
In view of the asymptotic  \cite{Lgp}
\begin{equation}
L_n^{(\alpha-n)}(z) \approx e^z \, {\alpha\choose n}, 
                    \hspace*{1.2cm}(n\rightarrow \infty)
\end{equation}
where the (generalized) binomial coefficients
\begin{equation}
{\alpha \choose k} := \frac{\alpha (\alpha-1) (\alpha-2)
                         \cdots (\alpha-k+1)}{k!}, 
\end{equation}
one finds from (\ref{cns})
\begin{equation}
\frac{c_{n+1}}{c_n} \sim \frac{1}{\kappa}\frac{\alpha-n}{n+1} 
          \rightarrow - \frac{1}{\kappa},  
                  \hspace*{1.8cm}(n\rightarrow\infty)
\end{equation}
whenever $\alpha\not\in \mathbb{N}$. Therefore, 
unless $\alpha$ is a nonnegative integer, 
the solution of eq. (\ref{cns}) is the {\em dominant} solution,
the power series solution (\ref{pss})
has only a {\em finite} radius of convergence and thus
does not belong to ${\cal B}$. 
Only if the condition (\ref{alqc}) is satisfied,  
$\alpha= x +\kappa^2$ is an {\em integer}, 
the {\em dominant} solution changes smoothly 
into the {\em minimal} solution of the recurrence 
(\ref{cns}), and the corresponding $F_{dho}$ vanishes.

In the case of the Rabi model, the recurrence (\ref{3trg}) becomes
\begin{equation}
c_{n+1} - \frac{f_n(x)}{(n+1)}\, c_n + \frac{1}{n+1}\, c_{n-1}=0,
\label{sa8}
\end{equation}
where $f_n(x)$ is given by 
\begin{equation}
f_n(x)=2\kappa+\frac{1}{2\kappa}
\left(n-x - \frac{\Delta^2}{n-x}\right),
\label{f-n}
\end{equation}
$\kappa=\lambda/\omega$, as in eq. (\ref{sa17}), and 
dimensionless $\Delta=\mu/\omega$ ({\em cf.} 
eq. (A8) of Schweber \cite{Schw}, which has mistyped 
sign in front of his $b_{n-1}$,
and eqs. (4) and (5) of \cite{Br}).
At first glance the recurrence (\ref{sa8}) 
does not  reduce to  (\ref{sa17})
for $\Delta=0$ as one would expect from (\ref{rabih}). 
The equivalence of eqs. (\ref{sa17}) and (\ref{sa8}) 
for $\Delta=0$ is disguised 
by the fact that the recurrence (\ref{sa8}) 
has been obtained after the unitary transformation of $\hat{H}_R$ 
in (\ref{rabih}) induced by
$D=\exp[\kappa(\hat{a}^\dagger - \hat{a})\sigma_1]$  \cite{Schw}.
Thereby the dimensionless energy parameter $x$ 
in eqs. (\ref{sa8}), (\ref{f-n}) is $x=(E/\omega)+\kappa^2$.

Both examples (\ref{sa17}) and (\ref{sa8})
represent the special Poincar\'{e} case characterized in that
the respective $a_n$ and $b_n$ in 
(\ref{rcd}) have finite limits $\bar{a}$ and $\bar{b}=0$ \cite{Schw,Gt,CP}.
The recurrences (\ref{sa17}) and (\ref{sa8}) correspond to the 
choice of $\delta=0$ and $\upsilon=-1$ in (\ref{rcd}), and hence $\tau=1$.
They only differ in the value of $\bar{a}=1/\kappa$ in (\ref{sa17}) and 
$\bar{a}=-1/(2\kappa)$ in (\ref{sa8}), whereas $\bar{b}=1$ in both
examples. Linearly independent solutions of (\ref{sa8}) can 
be written for $n\gg 1$ as
\begin{equation}
\nu_n = \left\{
\begin{array}{ll}
 1/(n \kappa^n), & \mbox{dominant}
\\
  n^{x}\kappa^n/\Gamma(n+1), & \mbox{minimal},
\end{array}\right.
\label{ldps}
\end{equation}
in agreement with (A.29a,b) of Schweber \cite{Schw}
and eq. (\ref{minas}) [note that $n^{x}$ leads to ${\cal O}(1)$ term
in (\ref{mins})]. 
In order to establish asymptotic of 
$\phi_m(z)$ for $z\rightarrow\infty$, one defines
\begin{equation}
\chi(z)=\sum_{n=0}^\infty \nu_n z^n,
\end{equation}
where the minimal solution (\ref{ldps}) has been substituted 
for $\nu_n$. Obviously, 
$\phi_m(z)\sim \chi(z)$ for $z\rightarrow\infty$.
The asymptotic of $\chi(z)$ can be
determined by the saddle point of the Euler-Maclaurin 
integral representation of $\chi(z)$ \cite{AMds,F,J}.
For $z\rightarrow\infty$ and
$|\arg z|\leq\pi/2$, which comprises the positive real axis
with the largest growth of $\chi(z)$,  one finds
a unique saddle point in the cut complex plane that 
completely governs the asymptotic behaviour 
of $\chi(z)$ \cite{AMds,F,J},
\begin{equation}
\chi(z) \sim  h(z,x) \, 
e^{\kappa z +  [x-(1/2)]\ln [\kappa z + x-(1/2)]} +
{\cal O}(|z|^p),
\label{emclra}
\end{equation}
where $h(z,x)=\sqrt{\kappa z + x - (1/2)}$ 
and $p$ is real constant satisfying $-1<p <0$.
As expected, $\chi(z)$ behaves essentially as $e^{\kappa z}$,
because the asymptotic of the minimal 
solution (\ref{ldps}) is only a 
slightly perturbed version of $1/\Gamma(n+1)$.
Consequently, $\phi_m(z)\in{\cal B}$ for all model parameters.
The leading order behaviour of the recurrence coefficients in (\ref{sa17})
for $n\rightarrow\infty$ resembles that of the Rabi model and
the above conclusions apply also to the case of 
the displaced harmonic oscillator.

\section{Discrete symmetries}
The Rabi Hamiltonian $\hat{H}_R$ [eq. (\ref{rabih})] is known
to possess a discrete $\mathbb{Z}_2$-symmetry corresponding 
to the constant of motion, or parity, 
$\hat{P}=\exp(i\pi \hat{J})$  \cite{Ks,Br}, where
\begin{equation}
\hat{J}=\hat{a}^\dagger \hat{a}+\frac{1}{2}\,(1+\sigma_3)
\end{equation}
is the familiar operator known to generate a continuous $U(1)$
symmetry of the JC model \cite{JC,Br}.
In contrast to ref. \cite{Br}, our approach does not necessitate
any active use of any underlying discrete symmetry. 
This has been demonstrated in a recent comment \cite{AMcm},
where the regular spectrum of the Rabi model 
has been reproduced as zeros
of a corresponding $F_{Rd}(x)$ based
on the recurrence (\ref{sa8}). 
Obviously it was not possible then to determine what is the 
parity of a state corresponding to a given zero of $F_{Rd}(x)$.

\begin{figure}
\onefigure[scale=0.7]{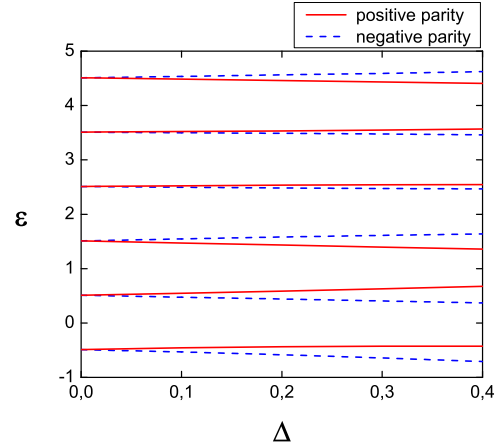}
\caption{Energy levels for $\kappa=0.7$ and $\omega=1$
as a function of $\Delta$ evolve from the doubly degenerate level 
corresponding to those of the displaced harmonic 
oscillator at $\epsilon_l=l-0.49$
to the parity eigenstates of the Rabi model.}
\label{fgddp}
\end{figure}

Nevertheless, not only any discrete symmetry 
can be easily incorporated 
in our approach, this can be accomplished 
more straightforwardly and more
easy than in the approach of Braak \cite{Br}.
Indeed, the Rabi Hamiltonian $\hat{H}_R$ 
is an example of a general 
Fulton and Gouterman Hamiltonian of a two-level system \cite{FGo}
\begin{equation}
\hat{H}_{FG}= A\,{\bf 1} + B\sigma_1 + C\sigma_3,
\label{hfg}
\end{equation}
with
\begin{equation}
A=\omega \hat{a}^\dagger \hat{a},~~~~~
B=\lambda (\hat{a}^\dagger + \hat{a}),~~~~~
C=\mu.
\label{abc}
\end{equation}
The Fulton and Gouterman symmetry operation $\hat{g}$ \cite{FGo} 
is realized by {\em reflections}
\begin{equation}
\hat{a}\rightarrow -\hat{a},~~~~~~ 
            \hat{a}^\dagger\rightarrow -\hat{a}^\dagger,
\label{fgrm}
\end{equation}
which leave the boson number operator 
$\hat{a}^\dagger \hat{a}$ invariant.
Because $[\hat{g},A]=[\hat{g},C]=\{ \hat{g},B\}=0$, 
$\hat{g}\sigma_3$ is 
the symmetry of $\hat{H}_R$ \cite{FGo}.
To any cyclic $\mathbb{Z}_2$ operator, such as 
$\hat{g}\sigma_3$, one can associate 
a pair of {\em projection operators}
\begin{equation}
P^\pm = \frac{1}{2}\, ( 1 \pm \hat{g}\sigma_3),~~~~~\left(P^\pm\right)^2 = P^\pm.
\label{fgpo}
\end{equation}
However, because of $\hat{g}\sigma_3$, the projectors $P^\pm$ do not mix
the upper and lower components of a wave function $\phi$
(conventional Pauli representation of $\sigma_1$ and $\sigma_3$ is assumed).
In order to employ the Fulton and Gouterman reduction in the 
{\em positive} and {\em negative parity spaces}, wherein one component 
of $\phi$ is generated from the other by means of the operator $\hat{g}$,
one is forced to work in a
unitary equivalent {\em single-mode spin-boson picture}
\begin{equation}
\hat{H}_R =
\omega \hat{a}^\dagger \hat{a} + \mu \sigma_1  
         + \lambda\sigma_3 (\hat{a}^\dagger + \hat{a}).
\end{equation}
The transformation is accomplished 
by means of the unitary operator $U=(\sigma_1 + \sigma_3)/\sqrt{2} = U^{-1}$.
The transformation interchanges the expressions for $B$ and $C$ in (\ref{abc}), 
resulting in $\hat{g}\sigma_1$
becoming the symmetry of $\hat{H}_R$. 
Eqs. (4.12) and (4.13) of \cite{FGo} then yield
\begin{equation}
 [ \omega \hat{a}^\dagger \hat{a}
   +  \lambda (\hat{a}^\dagger + \hat{a})  
          \pm \mu \hat{g}]\phi^\pm = E^\pm\phi^\pm,
\label{rbsp}
\end{equation}
where the superscripts $\pm$ denote the positive and negative
parity eigenstates of $P^\pm$ [with $\sigma_3$ being replaced
by $\sigma_1$ in (\ref{fgpo})].
Working in the Bargmann space,
\begin{equation}
 [\omega z\partial_z + \lambda (z+\partial_z) 
               \pm \mu \hat{g}]\phi^\pm = E^\pm\phi^\pm.
\label{rbmb}
\end{equation}
Eqs. (\ref{rbmb}) are equivalent to a coupled system of first-order eqs. (10) 
of the supplement to \cite{Br}.
\begin{figure}
\onefigure[scale=0.7]{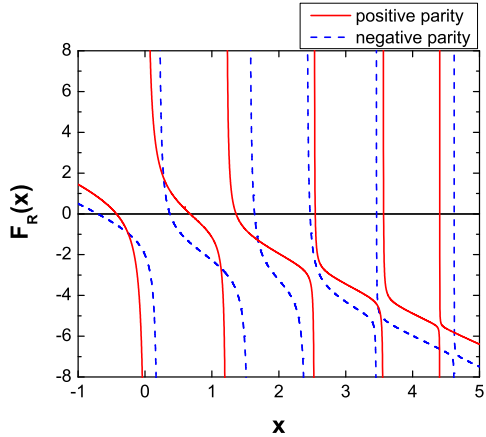}
\caption{$F_{R}(x)$ for $\kappa=0.7$, $\Delta=0.4$, and $\omega=1$,
{\em i.e.} the same parameters as in fig. 1 of
ref. \cite{Br}, shows zero at $\approx -0.707805$
for a negative parity state and at $\approx -0.4270437$ 
for a positive parity state.
After addition of $\kappa^2=0.49$ they correspond 
to the zeros at $-0.217805$ and
 $0.0629563$, respectively, in fig. 1 of refs. \cite{Br,AMcm}.}
\label{fgmrb}
\end{figure}
In contrast to \cite{Br}, one does not need any 
ill motivated substitution to arrive at the symmetry 
resolved spectrum of the Rabi model.
Assuming power series expansions 
$\phi^\pm (z) =\sum_{n=0}^\infty c_n^\pm z^n$ for 
the positive and negative parity
states,
one arrives directly at the following three-term recurrence
\begin{eqnarray}
 \lefteqn{
 c_{n+1}^\pm +\frac{1}{\kappa (n+1)}\, [n  - x  \pm(-1)^n\Delta]c_{n}^\pm 
}\hspace*{3cm}
\nonumber\\
      &&    + \frac{1}{n+1}\, c_{n-1}^\pm = 0,
\label{rbmb2}
\end{eqnarray}
where $x=E^\pm/\omega$ and $\kappa=\lambda/\omega$, 
as in eq. (\ref{sa17}), and
$\Delta=\mu/\omega$, as in eq. (\ref{sa8}).
In arriving at eqs. (\ref{rbmb2}) we have merely used that
\begin{equation}
\hat{g}\phi^\pm (z)=\phi^\pm (-z) =\sum_{n=0}^\infty (-1)^n c_n^\pm z^n.
\label{rbmbr}
\end{equation}
Upon defining corresponding $F^\pm_R(x)$ by means of
eqs. (\ref{fdf}) and (\ref{eth}), one can not only recover 
the regular spectrum of
the Rabi, but also distinguish different parity eigenstates.
This is demonstrated in fig. \ref{fgmrb}.
Fig. \ref{fgddp} demonstrates that on setting $\Delta=0$
the parity eigenvalues become doubly-degenerate eigenvalues 
corresponding to those of a displaced harmonic oscillator. 
That was to be expected,
because the three-term recurrence (\ref{rbmb2}) 
then reduces to the three-term recurrence (\ref{sa17}).

\section{Discussion}
\label{sec:disc}
In his recent letter \cite{Br}, Braak claimed to solve the Rabi model
analytically (see also Viewpoint by Solano \cite{Sln}). 
He suggested that a {\em regular} spectrum of the Rabi model was 
given by the zeros of transcendental functions $G_\pm(x)$
in the variable $x=(E/\omega)+\kappa^2$ \cite{Br},
\begin{equation}
G_\pm(x)=\sum_{n=0}^\infty K_n(x)
\left[1\mp\frac{\Delta}{x-n}\right]\kappa^n.
\label{sol}
\end{equation}  
Here the coefficients $K_n(x)$ were obtained recursively 
by solving the Poincar\'{e} difference equation (\ref{sa8}) upwardly
starting from the initial condition (\ref{rbc}).
Braak needed one page of arguments 
(between eq. (10) on p. 2 and the end of the first paragraph on p. 3
of the on-line supplement to \cite{Br})
to arrive at his $G_\pm(x)$. The arguments involved 
an ill motivated substitution
and that a {\em sufficient} condition for the vanishing 
of an analytic function $G_+(x; z)$ (defined 
by eq. (16) of his supplement) for {\em all} $z\in\mathbb{C}$
is if it vanishes at a {\em single} point $z=0$.
The argument is essential to arrive at (\ref{sol}). 
However, as an example of any homogeneous
polynomial shows, the argument is obviously invalid.
One needs additionally that $(d/dz)G_\pm(x; z)=0$ at $z=0$ \cite{MPS}.
At zeros of $G_\pm(x)$ the coefficients $K_n$'s in (\ref{sol})
have to be the minimal solutions.
The difficulty of calculating minimal solutions can be
illustrated by an attempt to calculate the Bessel functions of the first
kind $J_n(x)$ for fixed $x=1$ by an {\em upward} 
three-term recurrence from the 
initial values $J_0(x)$ and $J_1(x)$. 
It turns out that all digits calculated in single precision 
came out illusory already for $n\ge 7$ ({\em cf.} 
Table 1 of \cite{Gt}).

We have advanced an alternative approach which does not
require any explicit solution of any difference equation and which
applies to an entire class of models. 
A symmetry is not required to characterize 
a quantum model in terms of a suitable $F(x)$.
Contrary to ref. \cite{Br}, $F(x)$ can be obtained directly 
from the recurrence coefficients 
characterizing an eigenvalue equation of a given model. 
In the case of the Rabi model, nothing but the 
elementary relation (\ref{rbmbr}) has been employed in arriving 
from (\ref{rbmb}) to (\ref{rbmb2}), and the corresponding 
$F_R^\pm(x)$ were unambiguously determined from (\ref{rbmb2})
through eqs. (\ref{fdf}) and (\ref{eth}). 
The resulting transcendental functions $F_R^\pm(x)$
are {\em different} from
$G_\pm(x)$ considered by Braak \cite{Br}. 
With the exception
of the displaced harmonic oscillator and the JC model, the 
zeros of $F(x)$ have to be still determined {\em numerically}.
Therefore we disagree with Solano's view \cite{Sln} that a mere 
characterization of a model
by a function such as $G_\pm(x)$ of ref. \cite{Br}, or our $F(x)$, 
is tantamount to solving the model
analytically in a closed-form. 

Our results rejuvenate
the Schweber quantization criterion $r_0+a_0=0$ known in the case when $r_0$ 
remained to be expressed in terms of continued
fractions [{\em e.g.} eq. (\ref{rncf})] \cite{Schw}.
The Schweber criterion was deemed impractical and has not been employed to solve 
for the quantized energy levels of any quantum model \cite{Schw,Ks,KL,Br}.
The criterion was either regarded to require a numerical diagonalization 
in a truncated Hilbert space 
(pp. 4-5 of the on-line Supplement to ref. \cite{Br}),
or simply refuted as being identically valid {\em irrespective} of the
value of an energy parameter $x$
(see for instance p. 4 of the on-line Supplement to ref. \cite{Br}).
With the help of the Euler theorem, which transforms the continued
fractions into an infinite series, our approach turned
the Schweber quantization criterion into an efficient 
computational tool.

Our approach has been demonstrated on the examples 
of a displaced harmonic oscillator
and the Rabi model. However, the outlined approach would work for any of
the models of the class ${\cal R}$. For instance, the JC model 
and the single-mode spin-boson form of a 
generalized Rabi model introduced in  \cite{Br},
\begin{eqnarray}
H_{R\theta} = \omega \hat{a}^\dagger\hat{a} 
+\lambda \sigma_3(\hat{a}^\dagger + \hat{a})+\theta \sigma_3 +\mu\sigma_1,
\label{hamz2}
\end{eqnarray}
where $\theta$ is a deformation parameter. 
Another example is a modified Rabi model with the interaction Hamiltonian 
$\hat{H}_{int} = i \hbar \lambda \sigma_1 (\hat{a}^\dagger - \hat{a})$.
The latter arises if the dipole interaction with a 
Fabry-P\'{e}rot cavity light mode is 
replaced by the interaction with a single-mode {\em plane-wave} field.

We have characterized our class ${\cal R}$ of models implicitly in that
their eigenvalue equation can be reduced to a {\em difference equation}
(\ref{3trg}) with its coefficients satisfying eq. (\ref{rcd}).
An interesting problem is to provide an explicit 
characterization of the models.
Eventually note that whereas the condition 
$\tau\geq 1/2$ ensures $\phi_m(z)\in{\cal B}$,
one needs only $\tau>0$ to define $F(x)$ by eq. (\ref{fdf})
if $m_0\ne 0$.

\section{Conclusions}
\label{sec:conc}
A general formalism has been developed 
which allows  to determine the spectrum 
of an entire class of quantum models as zeros of a 
corresponding transcendental function 
$F(x)$. The function can be analytically determined as an infinite series 
defined solely in terms of recurrence coefficients.
The class of quantum models comprises the displaced harmonic oscillator,
the Jaynes-Cummings (JC) model, 
the Rabi model, and a generalized Rabi model.
Applications of the Rabi model range from quantum optics and magnetic
resonance to solid state and molecular physics. The model
plays a prominent role in cavity QED and circuit QED,
and can be experimentally realized  in Cooper-pair boxes, flux
q-bits, in Josephson junctions or using trapped ions,
and is of importance for various approaches to quantum
computing. Therefore, our results could have 
implications for further theoretical and experimental work 
that explores the interaction between light and matter, 
from weak to strong interactions.
The ease in obtaining the spectrum is of importance 
regarding recent experimental advances in preparing 
({\em ultra})strongly interacting 
quantum systems, which can no longer be reliably described by the 
exactly solvable JC model. 
The relevant computer code has been made freely available online \cite{AMr}.

\acknowledgments

I thank Professor Gautschi for a discussion.
Continuous support of MAKM is largely acknowledged.


\end{document}